\documentclass[12pt,reqno]{amsart}
\usepackage{appendix}
\usepackage{graphics,graphicx,amsmath,amssymb,epsfig,stmaryrd,color,flafter,xspace,lscape,threeparttable,pdfpages}
\usepackage[round]{natbib}
\usepackage{csquotes,amsfonts,amsthm,geometry,array,booktabs,latexsym,caption,subcaption,multirow,tabularx,rotating}
\usepackage[font=small,labelfont=bf]{caption}
\usepackage{changepage,enumitem,mathrsfs}
\usepackage{amsmath, upgreek,tipa}
\usepackage{bbold}
\DeclareCaptionFont{tiny}{\tiny}
\everymath{\displaystyle}
\RequirePackage{setspace,indentfirst,epsfig,psfrag,ifthen,ifpdf}

\newtheorem{theorem}{Theorem}
\theoremstyle{Plain}

\newtheorem{assumption}{Assumption}
\newtheorem{subassumption}{Assumption\normalfont}[assumption]

\newenvironment{assumption*}
{\ifnum\value{subassumption}=0 \stepcounter{assumption}\fi\subassumption}
{\endsubassumption}
\newenvironment{assumption+}[1]
{\renewcommand{\thesubassumption}{#1}\subassumption}
{\endsubassumption}
\theoremstyle{definition}
\newtheorem{assump}{Assumption}

\newtheorem*{definition*}{Definition}

\newtheorem{lemma}{Lemma}

\numberwithin{equation}{section}
\begin{document}
	\title{Changes-In-Changes For Discrete Treatment}
	\author{Onil Boussim}
	\thanks{ }
	\address{Penn State}
	
	
	
	\begin{abstract}
	This paper generalizes the changes-in-changes (CIC) model to handle discrete treatments with more than two categories, extending the binary case of \cite{athey2006}. While the original CIC model is well-suited for binary treatments, it cannot accommodate multi-category discrete treatments often found in economic and policy settings. Although recent work has extended CIC to continuous treatments, there remains a gap for multi-category discrete treatments. I introduce a generalized CIC model that adapts the rank invariance assumption to multiple treatment levels, allowing for robust modeling while capturing the distinct effects of varying treatment intensities.

		\vskip20pt
		

		\vskip10pt
		
	\end{abstract}
	
	\maketitle
	
	\newpage
	
	
\section{Introduction}

In a seminal paper, \cite{athey2006} introduced the changes-in-changes (CIC) model as an alternative to the widely used difference-in-difference (DID) method to evaluate the effects of policy changes. Unlike the standard DID approach, which primarily estimates average treatment effects, the CIC model aims to identify the entire potential outcome distribution for the treatment group had they not received the treatment and for the untreated group had they received the treatment. This nonlinear model provides a more comprehensive understanding of treatment effects by addressing unobserved heterogeneity and allowing for more nuanced insights into the distributional impacts of interventions.
The CIC model relies on specific assumptions to achieve its identification results, focusing on binary treatments. However, many real-world applications involve treatments that are discrete and non-binary, such as different levels of education, various job training programs, or multiple intensities of health interventions. The original CIC framework's binary treatment limitation constrains its applicability in these contexts.

Recent advancements, such as those by  \cite{d2023nonparametric}, have extended the CIC model to continuous treatments. Their approach generalizes the main assumptions of the CIC model and introduces a crossing point assumption applicable only to continuous treatments. Although this represents significant progress, it remains unsuitable for ordered or unordered discrete treatments, leaving a gap in the literature for handling multicategory discrete treatments effectively.
Despite these advancements, there is a lack of point identification results for discrete, ordered, or unordered treatments. This paper addresses this gap by proposing a set of identifying conditions for different parameters in the discrete treatment case, thus generalizing the results of the CIC model to accommodate ordered and unordered discrete treatments. 

Our extended CIC model adapts the original framework to handle multiple discrete treatment levels, preserving the robustness of the rank invariance assumption while accommodating the complexities of multi-category treatments. We develop a methodological framework that allows for the inclusion of covariates and the estimation of heterogeneous treatment effects across different subpopulations.

\section{Analytical Framework and Identification}

\subsection{General CiC Model}

For a possibly random variable  $Y$, $F_{Y}$ denotes the distribution of $Y$ and $F^{-1}_{Y}$ its corresponding quantile function. I  use calligraphic letters to denote the support of the random variables, for example, $\mathcal{Y}$ denotes the support of $Y$. 

Let $Y \in \mathbb R$ be a continuous outcome. I consider two periods $(t=0, t=1)$.  Some fraction of the population is assigned to a discrete treatment $D \in \mathcal{D} = \{ d_{1}, d_{2},..., d_{m} \}, (d_{1}=0)$ between $t=0$ and $t=1$. $Y_{t d}$ is the potential outcome at time $t$ when treatment is $d \in \mathcal{D}$. Extending \cite{abadie2005semiparametric}, I consider the following model :
\[\left\{
	\begin{array}{c}
            Y_{0} = Y_{00}\\
  		Y_{1} = \sum_{i=1}^{m} \mathbb 1\{D=d_{i} \} Y_{1d} 
	\end{array}
	\right.
\]

where $Y_{t}$ denotes the observed outcome at period $t$ and $Y_{td}$ denotes the potential outcome in period $t \in \{0,1\}$ and treatment status $d \in \mathcal{D} = \{ d_{1}, d_{2},..., d_{m} \}$, $d_{1}=0$. Let $p_{d} =  \mathbb P(D = d)$. The goal of this paper is to provide conditions for the identification of the counterfactual distributions:  $F_{Y_{1d}}(y)$ and $F_{Y_{1d} \vert D=d'}(y) \equiv \mathbb P(Y_{1d} \leq y  \vert D = d')$. Unlike the binary case, in the discrete case, many parameters can be of interest:

\[ QTE (\tau, d, d' ) = F^{-1}_{Y_{1d}}(\tau) - F^{-1}_{Y_{1d'}}(\tau) \]
\[ ATE  (d, d') = \int_{0}^{1}F^{-1}_{Y_{1d}}(\tau)d\tau - \int_{0}^{1} F^{-1}_{Y_{1d'}}(\tau)d\tau \]
\[ QTT (\tau,d, d' \vert d'') = F^{-1}_{Y_{1d} \vert D=d''}(\tau) - F^{-1}_{Y_{1d'} \vert D=d''}(\tau) \]
\[ ATT (d, d' \vert d'' ) =  \mathbb E(Y_{1d}-Y_{1d'} \vert D=d'') \]
And for ordered treatment, 
\[ ACRT (d_{j} \vert d) = ATT (d_{j}, d_{j-1}\vert d)  \]
\[ ACR (d_{j}) = ATE  (d_{j}, d_{j-1})  \]

where $QTE (\tau, d, d )$ is the $\tau$-quantile effect of treatment $d$ compared to treatment $d$  and $QTT(\tau,d \vert d)$ is the $\tau$-quantile effect of treatment $d$ compared to treatment $d$ in the post-treatment period on units that experienced dose $d$.

\subsection{Identification with weak rank stability}

For identification, I consider a set of assumptions similar  \cite{athey2006} and a generalization of the condition on the ranks. Therefore, I consider the following  identification assumption:

\begin{assumption}[Strict Monotonicity]\label{ass3}
\,\mbox{}\, \\
For each $d \in \mathcal{D}$, there exist two strictly increasing functions $h_{t}(.), t\in \{0,1\}$ and two uniformly distributed random variables over $[0, 1]$ $U_{td}$ such that $Y_{td} = h_{t}(U_{td})$.
\end{assumption}

\begin{assumption}[Weak Rank Stability]\label{ass3}
\,\mbox{}\, \\
$$U_{00} \vert D=d \sim U_{10}\vert D = d $$
\end{assumption}

\begin{assumption}[Support Condition]\label{ass3}
\,\mbox{}\, \\
$$ \mathcal{Y}_{t0 \vert d}  \subseteq  \mathcal{Y}_{t0\vert 0} $$
\end{assumption}

All these assumptions are virtually the same found in \citealp{athey2006} generalized to multiple treatment arms. Assumption 1 is similar to Assumption 3.1-3.2 of \cite{athey2006}. It is guaranteed for continuous random variables by the Skorohod quantile representation. Assumption 2 is close to Assumption 3.3 in \cite{athey2006}. It requires that the distribution of potential outcomes for individuals in a given group remains unchanged over time in the absence of treatment. This assumption ensures that any observed changes in the distribution of outcomes over time for the treated group can be attributed to the treatment rather than other time-varying factors. This helps in isolating the causal effect of the treatment by comparing the changes in the treated group with the changes in the control group. Assumption 3 is similar to the support condition (Assumption 3.4) in \cite{athey2006}. It means that the support of the untreated potential outcome of the group $D=d$ is included in the support of the untreated potential outcome of the group $D=0$.

These three assumptions together generalize to the discrete treatment case the identifying assumptions in the binary treatment changes in the changes model of \cite{athey2006}. Before discussing the identification, I derive this intermediate result useful for the identification proof but that also gives an alternative interpretation of the model. From Sklar theorem, we know that there exists a unique subcopula C such that:
$$ \textit{ For all y} \in \mathcal{Y}, $$
\begin{eqnarray*}
    \mathbb P(Y_{t0} \leq y,   D = d) = F_{Y_{t0}, D}(y,d)
    = C_{Y_{t0}, D}(F_{Y_{t0}} (y), p_{d})
\end{eqnarray*}

Under These assumptions, we can obtain the following lemma :
\begin{lemma}[Equivalence 1]
$$ \textit{Assumption 1, 2  and 3 together are equivalent to :} $$ 
$$C_{Y_{00},D}(u,p_{d}) = C_{Y_{10},D}(u,p_{d})$$ and $u \rightarrow C_{Y_{t0},D}(u,p_{d})$ is strictly increasing in $u$
\end{lemma}
This result was first established by \cite{ghanem2023evaluating} for binary treatment. I generalize it to discrete treatment. It requires that the dependence structure between the distribution of the potential untreated outcome and the membership of the group (or the assignment of treatment) remains consistent over time. One of the primary advantages of having a copula restriction is its invariance property. This copula invariance property ensures that our main identification result remains unaffected by any strictly monotonic transformation. Here is my first main identification result :

\begin{theorem}[Identification]
$$ \textit{Under Assumption 1-3 hold, for each } y \in \mathcal{Y} \textit{ we have .}$$ 
\[  F_{Y_{10} \vert D=d} (y) =F_{Y_{0}\vert D=d} \left(Q_{Y_{0} \vert D=0} \left(F_{Y_{1}\vert D=0}(y) \right) \right)\]
Therefore we can only identify the following:
\[ QTT (\tau,d,0 \vert d) = F^{-1}_{Y_{1d} \vert D=d}(\tau) - F^{-1}_{Y_{10} \vert D=d}(\tau) \]
\[ ATT (d, 0 \vert d ) = ATT (d \vert d ) =  \mathbb E(Y_{1d}\vert D=d) -  \int_{0}^{1}F^{-1}_{Y_{1d_{0}} \vert D=d} \]
\end{theorem}

Theorem 1 generalizes to the discrete treatment of the result in the binary case of \cite{athey2006}. Likewise, it does not impose any restriction on the heterogeneity of potential outcomes within a period or between periods. We specifically do not impose restrictions on the evolution of the distribution of the untreated potential outcome across time. The intuition behind the identification result is very simple and can be summarized as follows: In the first period, we observe the joint distribution $\mathbb P(Y_{00} \leq y, D=0)$ and both marginal distributions $\mathbb P(Y_{00} \leq y)$ and $\mathbb P(D=d)$. We can recover the copula   $C_{Y_{00},D}(u, p_{d})$ in the first period. Then, since we assume the dependence structure to be stationary over time and observe the joint distribution $\mathbb P(Y_{10} \leq y, D=0)$ in the second period, we can recover the marginal distribution $\mathbb P(Y_{10} \leq y)$ and also $\mathbb P(Y_{10} \leq y \vert D=d)$.

Theorem 2 proposes a more robust way to identify $ATT (d, 0 \vert d ) = ATT (d\vert d )= \mathbb E(Y_{1d} - Y_{10} \vert D=d)$ compared to the parallel trends assumption of \cite{callaway2024difference}.
They assume that :
\[ \mathbb E (Y_{10} - Y_{00} \vert D= d) = \mathbb E (Y_{10} - Y_{00} \vert D= 0) \]
This assumption says that the average evolution of the outcomes that the units with any treatment $d$ would have experienced without treatment is the same as the evolution of outcomes that units in the untreated group experienced. Like the natural parallel trend assumption, this one is also not invariant to a monotonic transformation since it restricts the covariance, and the covariance is not invariant to a monotonic transformation. It also restricts the evolution of the marginal distribution of $Y_{t0}$ over time and the dependence between $Y_{t0}$ and $D$.

Unlike the parallel trends assumption, the copula stability assumption is invariant to monotonic transformation and does not constrain the evolution of the marginal distribution over time, yet it relies only on the stability of the horizontal copula that governs the relationship between $Y_{t0}$ and $D$.

Despite the advantages of weak rank stability, many parameters of interest in this case cannot be recovered, therefore, I introduce a stronger assumption in the following section.

\subsection{Identification With Strong Rank Stability}

Since we cannot identify some parameters like for example average causal responses (ACR) under the dependence stability assumption, this suggests that learning about this type of parameter requires stronger assumptions. These assumptions are close to those found in the continuous CiC model of \cite{d2023nonparametric}.

\begin{assumption}[Strong Rank Stability]\label{ass3}
\,\mbox{}\, \\
$$U_{0d} \vert D=d' \sim U_{1d}\vert D = d' $$
\end{assumption}

\begin{assumption}[Support Condition]\label{ass3}
\,\mbox{}\, \\
$$ \mathcal{Y}_{td \vert d}  \subseteq  \mathcal{Y}_{td'\vert d} $$
\end{assumption}

Assumptions 1 and 4 together are similar to Assumption 1 of \cite{d2023nonparametric}. and Assumption is also a support condition we need. Like previously these assumptions can be stated in terms of a copula Invariance condition. 

It requires that the distribution of potential outcomes for individuals in a given group remains unchanged over time in the presence of any other treatment arm.

We have from Sklar's theorem $\mathbb P(Y_{td} \leq y, D = d') =  C_{Y_{td},D}(F_{Y_{td}}(y),p_{d'}) $

\begin{lemma}[Equivalence 2]
$$ \textit{Assumption 1, 4  and 5 together are equivalent to :} $$ 
$$ \textit{For all d,d', and all u} \in [0,1], $$ 
$$ C_{Y_{0d},D}(u,p_{d'}) = C_{Y_{1d},D}(u,p_{d'})$$
\end{lemma}

Assumption 3 necessitates that the dependence structure between the distribution of the potential outcome under treatment $d$ and group membership (or treatment assignment) remains consistent over time. This assumption can also be seen as some generalization of the traditional CiC assumptions. Here is the statement of the identification result. 

\begin{theorem}[Identification]
$$ \textit{Let Assumption 1, 4 and 5 hold, for each } y \in \mathcal{Y} \textit{ we have .}$$ 

\[  F_{Y_{1d}} (y) = F_{Y_{0}}(Q_{Y_{0} \vert D=d}(F_{Y_{1}\vert D=d}(y)))\]
\[  F_{Y_{1d} \vert D=d'} (y) =F_{Y_{0}\vert D=d'} \left(Q_{Y_{0} \vert D=0} \left(F_{Y_{1}\vert D=0}(y) \right) \right)\]

Therefore, we can identify the following:
\[ QTE (\tau, d, d' ) = F^{-1}_{Y_{1d}}(\tau) - F^{-1}_{Y_{1d'}}(\tau) \]
\[ ATE  (d, d') = \int_{0}^{1}F^{-1}_{Y_{1d}}(\tau)d\tau - \int_{0}^{1} F^{-1}_{Y_{1d'}}(\tau)d\tau \]
\[ QTT (\tau,d, d' \vert d'') = F^{-1}_{Y_{1d} \vert D=d''}(\tau) - F^{-1}_{Y_{1d'} \vert D=d''}(\tau) \]
\[ ATT (d, d' \vert d'' ) =  \mathbb E(Y_{1d}-Y_{1d'} \vert D=d'') \]
And for ordered treatment, 
\[ ACRT (d_{j} \vert d_{j}) = ATT (d_{j}, d_{j-1}\vert d_{j})  \]
\[ ACR (d_{j}) = ATE  (d_{j}, d_{j-1})  \]
\end{theorem}

\section{Applications  }

\section{Conclusion}

	\newpage
	\label{app:proofs}
	\appendix
	\label{app:proofs}
\section{Proofs of the results in the main text}

 \subsection{Proof of Lemma 1} 

From Sklar theorem, we know that there exists a unique subcopula C such that:
$$ \textit{ For all y} \in \mathcal{Y}, $$
\begin{eqnarray*}
    \mathbb P(Y_{t0} \leq y,   D = d) = F_{Y_{t0}, D}(y,d)
    = C_{Y_{t0}, D}(F_{Y_{t0}} (y), p_{d})
\end{eqnarray*}
We need to prove the equivalence between these 2 statements:
\begin{itemize}
\item  (Assumption 1-3) There exist two strictly increasing functions $h_{t}(.), t\in \{0,1\}$ and two uniformly distributed random variables over $[0, 1]$ $U_{t0}$ such that $Y_{t0} = h_{t}(U_{t0})$ and $U_{00} \vert D=d \sim U_{10}\vert D = d $
\item $C_{Y_{00},D}(u,p_{d}) = C_{Y_{10},D}(u,p_{d})$
\end{itemize}

Let  $p_{d} = \mathbb P(D=d)$. Since the cdf $F_{Y_{t0}}$ is continuous and strictly increasing, we have :
\[ Y_{t0} = Q_{Y_{t0}}(U_{t0}) \equiv h_{t}(U_{t0}) \]
By definition, $h_{t}$ is continuous and strictly increasing as is the quantile function $Q_{Y_{t0}}$.
\begin{eqnarray*}
    C_{Y_{00},D}(u,p_{d}) = C_{Y_{10},D}(u,p_{d})&\Leftrightarrow& C_{U_{00},D}(u,p_{d}) = C_{U_{10},D}(u,p_{d})\\
    &\Leftrightarrow& 
    \mathbb P(U_{00} \leq u, D = d) = \mathbb P(U_{10} \leq u, D = d)\\
    &\Leftrightarrow& 
   \frac{\mathbb P(U_{00} \leq u, D = d)}{\mathbb P(D=d)} = \frac{\mathbb P(U_{10} \leq u, D = d)}{\mathbb P(D=d)}\\
    &\Leftrightarrow& 
    \mathbb P(U_{00} \leq u \vert D = d) = \mathbb P(U_{10} \leq u \vert  D = d)
\end{eqnarray*}
The first equivalence follows from the invariance principle, and the second holds from Sklar’s theorem.
$$ \textit{The function } u \mapsto C_{Y_{t0},D}(u,p_{d}) \textit{ is strictly increasing. In fact, we know that :}$$ 
\[ F_{Y,D} (Q_{Y_{t0}}(u), d) = C_{Y_{t0},D}(u,p_{d}) \]
Let $ u, u' \in [0,1]$. We have 
\begin{eqnarray*}
    u < u' &\Rightarrow&  Q_{Y_{t0}}(u) < Q_{Y_{t0}}(u') \\&\Rightarrow& F_{Y_{t},D}(Q_{Y_{t0}}(u), d) <  F_{Y_{t},D}(Q_{Y_{t0}}(u'), d)\\
    &\Rightarrow&
    C_{Y_{t0},D}(u,p_{d}) < C_{Y_{t0},D}(u',p_{d})
\end{eqnarray*}

\subsection{Proof of Theorem 1} 
Now, I derive the following. Fix $y \in \mathcal{Y}$. We know that:
\begin{eqnarray*}
    F_{Y_{1} \vert D=0}(y) 
    &=& F_{Y_{0} \vert D=0}(Q_{Y_{0} \vert D=0}(F_{Y_{1} \vert D=0}(y))) \\
    F_{Y_{1}, D}(y, 0) 
    &=& F_{Y_{0} , D}(Q_{Y_{0} \vert D=0}(F_{Y_{1} \vert D=d}(y)), 0) \\
    C_{Y_{10},D}(F_{Y_{10}}(y), p_{0})  &=&  C_{Y_{0},D}( F_{Y_{0}}(Q_{Y_{0} \vert D=0}(F_{Y_{1} \vert D=0}(y))   , p_{0})\\
    C_{Y_{00},D}(F_{Y_{10}}(y), p_{0})  &=&  C_{Y_{0},D}( F_{Y_{0}}(Q_{Y_{0} \vert D=0}(F_{Y_{1} \vert D=0}(y))   , p_{0})\\
    C_{Y_{0},D}(F_{Y_{10}}(y), p_{0})  &=&  C_{Y_{0},D}( F_{Y_{0}}(Q_{Y_{0} \vert D=0}(F_{Y_{1} \vert D=0}(y))   , p_{0})\\
    F_{Y_{10}}(y) &=& F_{Y_{0}}(Q_{Y_{0} \vert D=0}(F_{Y_{1} \vert D=0}(y))
\end{eqnarray*}

The passage from the third to the fourth line requires the use of Lemma 1. And the passage from line 5 to line 6 requires the use of lemma 2. Now we also know that:
\begin{eqnarray*}
\mathbb P(Y_{10} \leq y \vert D=d) &=& \frac{ \mathbb P(Y_{10} \leq y, D=d)}{\mathbb  P(D=d)}\\
&=&\frac{ C_{Y_{10},D}(F_{Y_{10}}(y), p_{d}) }{\mathbb  P(D=d)}\\
&=&\frac{ C_{Y_{0},D}(F_{Y_{10}}(y), p_{d}) }{\mathbb  P(D=d)}\\
&=&\frac{ \mathbb P(Y_{0} \leq Q_{Y_{0}}(F_{Y_{10}}(y)) , D=d)}{\mathbb  P(D=d)}\\
&=&\frac{ \mathbb P(Y_{0} \leq Q_{Y_{0} \vert D=0}(F_{Y_{1} \vert D=0}(y)) , D=d)}{\mathbb  P(D=d)}\\
&=&F_{Y_{0} \vert D=d}(Q_{Y_{0} \vert D=0}(F_{Y_{1} \vert D=0}(y)))
\end{eqnarray*}

\subsection{Proof of lemma 2}

From Sklar theorem, we know that there exists a unique subcopula C such that:
$$ \textit{ For all y} \in \mathcal{Y}, $$
\begin{eqnarray*}
    \mathbb P(Y_{td} \leq y,   D = d') = F_{Y_{td}, D}(y,d')
    = C_{Y_{td}, D}(F_{Y_{td}} (y), p_{d'})
\end{eqnarray*}
We want to prove that the 2 statements are equivalent  when $D$ is discrete: $p_{d'} = \mathbb P(D=d')$

$$ \textit{The two following statements are equivalent} $$ 
\begin{itemize}
\item  (Assumption 1,4) There exist two strictly increasing functions $h_{t}(.), t\in \{0,1\}$ and two uniformly distributed random variables over $[0, 1]$ $U_{td}$ such that $Y_{td} = h_{t}(U_{td})$ and $U_{0d} \vert D=d' \sim U_{1d}\vert D = d' $
\item $C_{Y_{0d},D}(u,p_{d'}) = C_{Y_{1d},D}(u,p_{d'})$
\end{itemize}

\begin{eqnarray*}
    C_{Y_{0d},D}(u,p_{d'}) = C_{Y_{1d},D}(u,p_{d'})&\Leftrightarrow& C_{U_{0d},D}(u,p_{d'}) = C_{U_{1d},D}(u,p_{d'})\\
    &\Leftrightarrow& 
    \mathbb P(U_{0d} \leq u, D = d') = \mathbb P(U_{1d} \leq u, D = d')\\
    &\Leftrightarrow& 
   \frac{\mathbb P(U_{0d} \leq u, D = d')}{\mathbb P(D=d')} = \frac{\mathbb P(U_{1d} \leq u, D = d')}{\mathbb P(D=d')}\\
    &\Leftrightarrow& 
    \mathbb P(U_{0d} \leq u \vert D = d') = \mathbb P(U_{1d} \leq u \vert  D = d')
\end{eqnarray*}
The first equivalence follows from the invariance principle, and the second holds from Sklar’s theorem.
 
\subsection{Proof of theorem 4}

Fix $y \in \mathcal{Y}$. We know that :

\begin{eqnarray*}
    F_{Y_{1} \vert D=d}(y) 
    &=& F_{Y_{0} \vert D=d}(Q_{Y_{0} \vert D=d}(F_{Y_{1} \vert D=d}(y))) \\
    F_{Y_{1}, D}(y, d) 
    &=& F_{Y_{0} , D}(Q_{Y_{0} \vert D=d}(F_{Y_{1} \vert D=d}(y)), d)\\
    C_{Y_{1d},D}(F_{Y_{1d}}(y), p_{d})  &=&  C_{Y_{0},D}( F_{Y_{0}}(Q_{Y_{0} \vert D=d}(F_{Y_{1} \vert D=d}(y))   , p_{d})\\
    C_{Y_{0d},D}(F_{Y_{1d}}(y), p_{d})  &=&  C_{Y_{0},D}( F_{Y_{0}}(Q_{Y_{0} \vert D=d}(F_{Y_{1} \vert D=d}(y))   , p_{d})\\
    C_{Y_{0},D}(F_{Y_{1d}}(y), p_{d})  &=&  C_{Y_{0},D}( F_{Y_{0}}(Q_{Y_{0} \vert D=d}(F_{Y_{1} \vert D=d}(y))   , p_{d})\\
    F_{Y_{1d}}(y) &=& F_{Y_{0}}(Q_{Y_{0} \vert D=d}(F_{Y_{1} \vert D=d}(y))
\end{eqnarray*}
The passage from the third to the fourth line requires the use of Assumption 1. And the passage from line 5 to line 6 requires the use of Assumption 2. Now, we know that :
\begin{eqnarray*}
\mathbb P(Y_{1d} \leq y \vert D=d') &=& \frac{ \mathbb P(Y_{1d} \leq y, D=d')}{\mathbb  P(D=d')}\\
&=&\frac{ C_{Y_{1d},D}(F_{Y_{1d}}(y), p_{d'}) }{\mathbb  P(D=d')}\\
&=&\frac{ C_{Y_{0},D}(F_{Y_{10}}(y), p_{d'}) }{\mathbb  P(D=d')}\\
&=&\frac{ \mathbb P(Y_{0} \leq Q_{Y_{0}}(F_{Y_{10}}(y)) , D=d')}{\mathbb  P(D=d')}\\
&=&\frac{ \mathbb P(Y_{0} \leq Q_{Y_{0} \vert D=d}(F_{Y_{1} \vert D=d}(y) , D=d')}{\mathbb  P(D=d')}\\
&=&F_{Y_{0} \vert D=d'}(Q_{Y_{0} \vert D=d}(F_{Y_{1} \vert D=d}(y)))
\end{eqnarray*}

 \label{app:}

	
	
	\bibliographystyle{abbrvnat}
	\bibliography{ref}

	
\end{document}